\input{aipcheck}

\documentclass[
    ,final            % use final for the camera ready runs
  ]
  {aipproc}

\layoutstyle{8x11single}

\usepackage[latin1]{inputenc}
\usepackage{amsmath,amsfonts,amssymb,amsthm}

\theoremstyle{definition}

\theoremstyle{remark}

\newcommand{\norm}[1]{\left\Vert {#1} \right\Vert}
\newcommand{\abs}[1]{\left\vert {#1} \right\vert}
\newcommand{\set}[1]{\left\{ {#1} \right\}}
\newcommand{\prt}[1]{\left( {#1} \right)}
\newcommand{\scal}[1]{\left< {#1} \right>}

\newcommand{\sR}{{\mathbb R}}
\newcommand{\sC}{{\mathbb C}}
\newcommand{\implie}{\ \ \Rightarrow\ \ }
\newcommand{\limplie}{\ \ \Longrightarrow\ \ }
\newcommand{\equi}{\ \ \Leftrightarrow\ \ }
\newcommand{\lequi}{\ \ \Longleftrightarrow\ \ }

\newcommand{\A}{\mathcal{A}}

\renewcommand{\H}{\mathcal{H}}
\newcommand{\C}{\mathcal{C}}

\DeclareMathOperator{\tr}{tr}

\begin{document}

\title{Towards a noncommutative version of Gravitation}

\author{Nicolas Franco}{
  address={Groupe d'Applications Mathématiques aux Sciences du Cosmos\\ University of Namur FUNDP - Département de Mathématiques\\ Rempart de la Vierge, 8 B-5000 Namur Belgium}
  ,email={nicolas.franco@math.fundp.ac.be}
}

\begin{abstract}
Alain Connes' noncommutative theory led to an interesting model including both Standard Model of particle physics and Euclidean Gravity. Nevertheless, an hyperbolic version of the gravitational part would be necessary to make physical predictions, but it is still under research. We shall present the difficulties to generalize the model from Riemannian to Lorentzian Geometry and discuss key ideas and current attempts.
\end{abstract}

\maketitle

\section[Noncommutative Geometry]{Noncommutative Geometry}

Noncommutative Geometry is a theory which could be seen as an {\it algebrization of geometry} or an {\it geometrization of algebra}. It provides a way to translate geometrical concepts into an algebraic framework. In this paper we will focus on the algebraic translation of Lorentzian geometry - and so the mathematical tools for gravitation - by using the notion of spectral triples developed by Alain Connes.\\

Let's begin by reviewing the bases of noncommutative geometry. This theory arises at first from the so-called Gel'fand transform
$$\bigvee : {\bf A} \rightarrow C(\Delta({\bf A})) : a \leadsto \hat a,\qquad\hat a(\chi) = \chi(a)$$
where $ {\bf A}$ is a commutative C*-Algebra and $\Delta({\bf A})$ its space of characters (non-zero morphisms).\\

The Gel'fand-Neumark theorem states:\\

{\it For any (unital) commutative C*-algebra {\bf A}, the Gel'fand transform is an isometric isomorphism between {\bf A} and \mbox{$C(\Delta({\bf A}))$}.}\\

This Gel'fand transform gives a  functor between the category of  locally compact ({\it compact}) Hausdorff spaces and the category of ({\it unital}) commutative C*-algebras. Gel'fand-Neumark theorem insures that there is no lost of information by considering commutative C*-algebras instead of geometrical manifolds. The noncommutative generalization of geometry comes by considering not only commutative C*-algebras, but the extension to noncommutative C*-algebras.\\

Many geometrical concepts can be translated into the algebraic formalism. This leads to a kind of dictionary, each geometrical tool having its algebraic counterpart:

\begin{center}\begin{tabular}{rcl}
{\bf Geometry} &  & {\bf Algebra} \\
\hline
Point & $\leftrightarrow$ & Caracter \\
Locally compact space & $\leftrightarrow$ & C*-algebra \\
Fiber bundle & $\leftrightarrow$ & Finite projective module \\
Complex variable & $\leftrightarrow$ & Operator \\
Real variable & $\leftrightarrow$ & Hermitian operator \\
Infinitesimal & $\leftrightarrow$ & Compact operator \\
Integral & $\leftrightarrow$ & Trace \\
& ... & 
\end{tabular}\end{center}

\section[Spectral Triples]{Spectral Triples}

We now comes to Alain Connes' theory by defining its main ingredient \cite{C94} \cite{Var}.\\

A {\bf Spectral Triple} \mbox{$(\A,\H,D)$} is given by:
\begin{itemize}
\item an Hilbert space $\H$
\item an involutive algebra $\A$ of bounded operators on $\H$ 
\item an self-adjoint operator $D$ on $\H$ with compact resolvent such that $$[D,a] \in \mathcal B(\H)\qquad \forall a \in \A\qquad\text{({\it first order condition})}$$
\end{itemize}

The choose of a particular spectral triple is equivalent to the choose of a particular space endowed with its geometry. Commutative continuous algebras $\A$ for example correspond to usual manifolds. Indeed, if we take a compact Riemannian spin manifold $(M,g)$, we can construct the following spectral triple:
\begin{itemize}
\item $\A = C^\infty(M)$
\item $\H =  L^2(S)$
\item $D = \gamma^a e_a^\mu \prt{ \partial_{\mu} + \frac{1}{4} \omega_\mu^{ab}\gamma_{ab}}\qquad\text{({\it Dirac operator})}$
\end{itemize}
where $S$ is the spinor bundle over $M$ and \mbox{$\omega_\mu^{ab}$} the spin-connection.\\

Such spectral triple allows us to translate the differential structure of $M$  in algebraic terms: 
$$df  \quad\sim\quad [D,f]$$
$$f_0 df_1 \cdot\cdot\cdot df_p  \quad\sim\quad f_0 [D,f_1] \cdot\cdot\cdot[D,f_p]$$

The condition $$\norm{[D,f]}  \leq 1  \equi \norm{df} \leq 1$$ is equivalent to a Lipschitz condition on $f$, and allows us to construct a notion of distance on a spectral triple:

$$ d(p,q) = \sup_{f\in\A}\set{\abs{f(p) - f(q)} : \norm{[D,f]} \leq 1}$$
This function is equivalent to the usual distance on a Riemannian manifold
$$ d(p,q) = \inf \int_\gamma ds\qquad\prt{\gamma \text{ path from } p \text{ to } q}$$ but without reference to any path. It can be extended to an arbitrary spectral triple by considering pure states instead of functions. This distance function is a way to recover the information on the metric in a completely algebraic framework, and leads to Connes' reconstruction theorem \cite{C96} :\\

{\it Consider an spectral triple \mbox{$\prt{\A,\H,D}$} whose algebra $\A$ is commutative. Then there exists a compact Riemannian spin manifold M whose spectral triple \mbox{$\prt{C^\infty(M), L^2(S),D}$} coincides with  \mbox{$\prt{\A,\H,D}$}.}\\

The current model from Alain Connes \cite{MC207} \cite{MC08} is a model unifying Euclidian Gravity and a classical Standard Model, by making a product of two spectral triples, one with a continuous algebra corresponding to the gravitational part, the other with a discrete algebra corresponding to the bosonic part of the standard model:

$$(\A,\H,D) = \underbrace{(\A_1,\H_1,D_1)}_{\frac{\text{continuous algebra}}{\text{gravitational part}}} \otimes\underbrace{(\A_2,\H_2,D_2)}_{\frac{\text{discrete algebra}}{\text{standard model}}}$$

with $\A_2 = \sC \oplus \mathbb H \oplus M_3(\sC)$.\\

A fundamental action functional (spectral action) is defined
$$S =\tr\prt{ f\prt{\frac{D}{\Lambda}}} + \scal{\psi, D \psi}$$
where $\Lambda$ is a cut off parameter, $f$ a positive suitable function and the last term being added to handle the fermionic part.\\

An adaptation of the product geometry with Lorentzian signature (more precisely with a signature not equal to the dimension) was proposed by John Barret in \cite{Barret}.

\section[Lorentzian Gravity]{Lorentzian Gravity}

Current Connes' model is based on a Riemannian manifold describing Euclidian Gravity. But of course this does not correspond  to any physical reality! In order to build a physical version of the theory and apply it to real physical problems, one must find a Lorentzian version of the theory, so to build a complete translation of Lorentzian Geometry into an algebraic framework. However, research in this domain is only at a beginning state, and a complete solution to adapt the model to hyperbolic spaces is still out of sight. In the remaining of this paper, we will discuss about several problems that arise in this theory when dealing with Lorentzian spaces and the first attempts to solve them.\\

We will focus mainly on the construction of a Lorentzian distance formula similar to the Riemannian one:
$$ d(x,y) = \sup_{f\in\A}\set{\abs{f(x) - f(y)} : \norm{[D,f]} \leq 1}$$
The construction of such function can be seen as a 4-pieces puzzle, where each piece represents a difficult and still unsolved or partially solved problem:
\begin{itemize}
\item A spectral triple for gravity is based on a Hilbert space $\H = L^2(S)$ endowed with the positive definite inner product $(\psi,\phi) = \int_M \psi^*\phi\, d\mu_g$. But in Lorentzian geometry, this product is not positive definite any more, and the space is now a space with an indefinite inner product, with no norm available. A new space with a new kind of structure must be set.
\item The Dirac operator - carrying the information on the metric - is not an elliptic operator any more. The Lipschitz condition $\norm{[D,f]} \leq 1$ necessary for the distance function is not correctly defined. So one or more replacement conditions must be found, maybe by the use of a new operator more appropriate.
\item Causality is a new element that was absent in the Riemannian framework. Information on causality much be translated into the algebraic formalism, either by the use of an operator or by any external manner.
\item The old distance formula 
$$ d(x,y) = \sup_{f\in\A}\set{\abs{f(x) - f(y)} : \norm{[D,f]} \leq 1}$$
is clearly Riemannian in his form and must be adapted to Lorentzian criteria.
\end{itemize}

\section[Indefinite inner product]{Indefinite inner product space}

On the Lorentzian case, the space $\H = L^2(S)$ is made with the indefinite inner product $(\psi,\phi) = \int_M \psi^*\phi\, d\mu_g$. This product allows us to construct what is called a {\it Krein space} \cite{Bog}. The main idea of a Krein space is to define a fundamental symmetry on spectral triples which will have a similar effect than a Wick rotation to a metric.\\

Our space $\H$ can be split as the direct sum of complete orthogonal spaces $\H^+$ and $\H^-$ such that the inner product is positive definite on $\H^+$ and negative definite on $\H^-$. Thus the operator $\mathcal J = \text{id} \oplus -\text{id}$ defines a positive definite inner product by
$$\scal{\cdot,\cdot}_{\mathcal J} = (\cdot,\mathcal J\cdot).$$
This is our fundamental symmetry. So it is possible to redefine a spectral triple based on a Krein space \cite{Stro}. What is still unknown is the conditions we must imposed on $J$ to guarantee unicity.\\

\section[Dirac operator]{Dirac operator}

On the Riemannian case, the Dirac operator $D$ is an elliptic operator which carries the whole information about the dynamic. The norm of its commutator gives a Lipschitz condition:

$$\norm{[D,f]} \leq 1\lequi \text{$f$ Lipschitz}  $$

This condition is made by the use of a normed space equiped with a Cauchy-Schwarz inequality, which are both absent in the Lorentzian case.\\

So we should try to reconstruct a kind of Lipschitz condition by the use of Krein spaces. This approach is really non trivial because of the absence of Cauchy-Schwarz inequality in those spaces. Ongoing research in this field will be presented in a forthcoming paper.\\

A first attempt of generalization by the use of the Laplace-Beltrami operator can be found in \cite{Mor}.

\section[Causality]{Causality}

Causality is an interesting point in the generalization of spectral triples theory, because it is a concept that is completely absent of the Riemannian case. So we have to create a way to translate this concept in a noncommutative framework.\\

It appears, since \cite{Mor}, that the concept of causal functions could be a interesting ingredient while dealing with causality. A causal function is a continuous function which is non-decreasing along every future-directed causal curve. But the way to characterize such functions in noncommutative spaces is unknown at this time.\\

A first line in this challenge has been written by the introduction of causal sets theory \cite{Bes}. If we take a partially ordered set $M$, we can define the set of continuous isotone functions $I(M)$ by the set of functions $f:M\rightarrow \sR$ which satisfy:
\begin{itemize}
\item $f$ is continous
\item $\forall x,y \in M,\quad x \leq y \implie f(x) \leq f(y)$
\end{itemize}

In some spaces, the knowledge of $I(M)$ is sufficient to recover all information on the order:
$$\forall x,y \in M,\quad x \leq y \equi \forall f \in I(M)\ \ f(x) \leq f(y)$$
Such spaces are called completely separated ordered spaces. A globally hyperbolic spacetime $M$ is one of these spaces, where in this case the set $I(M)$ corresponds to the set of causal functions.\\

In \cite{Bes}, a noncommutative generalization of the set of isotone functions $I(M)$ was presented.  It is done by a couple $(I,C)$ called I*-algebra and defined by:
\begin{itemize}
\item $C$ is an unital C*-algebra
\item $I$ is a closed convex cone of hermitian elements of $C$ containing the constants and stable by two defined meet and join operations:
$$ a \wedge b = \frac{a+b}{2} - \frac{\abs{a-b}}{2},\qquad a \vee b = \frac{a+b}{2} + \frac{\abs{a-b}}{2}$$
\item $\overline{span(I)} = C$
\end{itemize}

In the compact case, it is possible to construct a functor between the category of completely separated ordered spaces and the category of I*-algebras. A given abelian I*-algebra determines one and only one order on a manifold. So a line has been added in the noncommutative dictionary about order. Nevertheless, one must find some extra conditions to guaranty that this order corresponds to Lorentzian causal structure.

\section[Lorentzian Distance]{Lorentzian Distance}

Along this last section, we will try to discuss on the form of the Lorentzian distance formula. Let's have a look at the origin of this function. The Riemannian distance formula
$$ d(p,q) = \sup_{f\in\A}\set{\abs{f(p) - f(q)} : \norm{[D,f]} \leq 1}$$
comes from the following equivalence: 
$$\norm{[D,f]} \leq 1\lequi \text{$f$ is Lipschitz with constant 1}  $$
$$\lequi \forall x,y :\ \  \abs{f(x) - f(y)} \leq d(x,y)  $$
$$\limplie\  \sup_{f\in\A}\set{\abs{f(p) - f(q)} : \norm{[D,f]} \leq 1}\ \leq\ d(x,y)  $$

Equality is obtained just by using the usual distance function $f(z) = d(z,q)$:
$$\abs{f(p) - f(q)} = \abs{d(p,q) - d(q,q)} = d(p,q)$$
which is well Lipschitz thanks to the triangle inequality:
$$\forall x,y\ \abs{f(x) - f(y)} = \abs{d(x,q) - d(y,q)} \leq d(x,y)$$

In order to build a Lorentzian distance function, we must adapt this function to the conditions for a Lorentzian distance:
\begin{itemize}
\item $d(x,x) = 0$
\item $d(x,y) > 0 \implie d(y,x) = 0$
\item $\text{if }x \prec y \prec z, \text{ then }\ d(x,z) \geq d(x,y) + d(y,z)\qquad\text{\it(reverse triangle inequality)}$
\end{itemize}

There are several ingredients we can replace or adapt:
$$ d(p,q) = \underbrace{\sup_{f\in\A}}_{\text{supremum?}}\set{\underbrace{\abs{f(p) - f(q)}}_{\text{difference?}} : \underbrace{\norm{[\underbrace{D}_\text{operator?},f]}}_{\text{Lipschitz condition?}} \underbrace{\leq 1}_{\text{constraint?}}}  $$

From now, we will work with two hypotheses, which are the followings:
\begin{itemize}
\item Work hypothesis 1: There exists a way (by use of an operator) to reproduce a Lipschitz-like condition in a Lorentzian framework.
\item Work hypothesis 2:  The algebraic Lorentzian distance will only be defined on points/characters causally connected, so there must exist an algebraic translation of causally structure.
\end{itemize}
In fact, these hypotheses simply assume that we have resolved the three first parts of the 4-pieces puzzle.\\

What happens if we want to conserve the same Lipschitz condition that in the Riemannian case ? We should have a function on this form, with $d_\sC$ a distance function on $\sC$:
$$ d(p,q) = \sup_{f\in\A}\set{d_\sC\!\prt{f(p), f(q)} : \text{\it $f$ Lipschitz}} \quad \prt{\text{}p \prec q}  $$

First problem is that the usual distance function $f(z) = d(z,q)$ is not a Lipschitz function any more, because of the reverse triangle inequality:
$$x \prec y \prec q,\ \abs{d(x,q) - d(y,q)} \geq d(x,y)$$

Second problem is on the function $d_\sC$. Let $f$ be an arbitrary Lipschitz function and assume that $d(x,y) > 0$, then we have:
$$d_\sC\!\prt{f(x), f(y)} \leq  \ d(x,y)\ \ \prt{> 0} $$
By Lorentizan conditions:
$$d_\sC\!\prt{f(y), f(x)} \leq  \ d(y,x) = 0 $$
$$\limplie d_\sC\!\prt{f(y), f(x)} = 0 $$
Last equation shows that, if we have a function $d_\sC$ such that $d_\sC\!\prt{a, b} = d_\sC\!\prt{b, a}$, then the distance $d$ will be automatically a null function. So if we want to conserve a Lipschitz condition, the old function $d_\sC\!\prt{f(x), f(y)} = \abs{f(x)-f(y)}$ must be replaced by a non-symetric one.  Moreover in this case, we can see that all functions $f \in \A$ must give by their ranges an information on the causal structure. So $\A$ must be restrained to a set of causal functions or similar.\\

Let's now try a different way to generalize the distance function, by introducing the concept of {\it dilatation}:

$$ d(p,q) = \inf_{f\in\A}\set{\abs{f(p) - f(q)} : \text{\it $f$ dilatation}} \quad \prt{\text{}p \prec q} $$
$$\text{\it $f$ dilatation} \lequi \forall x,y :\ \  \abs{f(x) - f(y)} \geq d(x,y)  $$

The main idea is to replace a supremum with a upper bound by a infimum with an equivalent lower bound. There are several advantages to use a dilatation condition instead of a Lipschitz in the Lorentzian case:
\begin{itemize}
\item We have trivially that $\ \inf_{f\in\A}\set{\abs{f(p) - f(q)} : \text{\it $f$ dilatation}} \ \geq\ d(p,q)  $
\item There is no need of restriction to causal functions 
\item Such kind of constraint - with an lower bound - is more natural in space with indefinite inner product 
\item The usual distance function $f(z) = d(z,q)$ respects the dilatation condition on light cones 
\end{itemize}

So what remains is to construct an equality function. This can be done with the hypothesis of a globally hyperbolic space-time. In a case of the distance between two points $p$ and $q$, with $p \prec q$, we can set the following function:
$$ f(z) = d(z,\C) - d(\C,z)$$
where \mbox{$\C$} is a Cauchy surface that contains $p$ and such that \mbox{$d(\C,q) = d(p,q)$}, and$$d(z,\C) = \sup_{t\in\C} \ d(z,t)$$ designs the distance between the point $z$ and the Cauchy surface. \\

We can check that this function gives the request equality:
$$\abs{f(p)-f(q)} = |\underbrace{d(p,\C)}_{0} - \underbrace{d(\C,p)}_{0} - \underbrace{d(q,\C)}_{0} + \underbrace{d(\C,q)}_{d(p,q)}| = d(p,q)$$

We must check that this function is a dilatation. We will separate the proof in 3 different cases (with two similar), depending on the localization of the Cauchy surface.\\

Let $x,y$ such that $x \prec\C\prec y$, than:
$$\abs{f(x)-f(y)} = |\underbrace{d(x,\C)}_{>0} - \underbrace{d(\C,x)}_{0} - \underbrace{d(y,\C)}_{0} + \underbrace{d(\C,y)}_{>0}| $$
$$ \geq d(x,t) + d(t,y) = d(x,y)$$
 with $t$ at the intersection between $\C$ and the geodesic from $x$ and $y$ (the existence of this geodesic is guaranteed by global hyperbolicity).\\
 
Let $x,y$ such that  $\C\prec x\prec y$: $\ $ (the remaining case is similar)
$$\abs{f(x)-f(y)} = |\underbrace{d(x,\C)}_{0} - \underbrace{d(\C,x)}_{>0} - \underbrace{d(y,\C)}_{0} + \underbrace{d(\C,y)}_{>0}| $$
$$ = \underbrace{d(\C,y)}_{\geq d(t,y)} - \underbrace{d(\C,x)}_{=d(t,x)} = d(t,y) - d(t,x) \geq d(x,y)$$
 with $t\in\C$ such that $d(t,x) = d(\C,x)$ (guaranteed by global hyperbolicity) and by using the reverse triangle inequality $d(t,x) + d(x,y) \leq d(t,y)$.

\section[Conclusion]{Conclusion}

We have seen that the generalization of Connes' theory to the hyperbolic case can be divided as a 4-pieces puzzle. There are attempts to solve each of these pieces, but no one is completely resolved at this time, nor the correct way to put all the pieces together. Moreover, to the 4-pieces puzzle we must also add the generalization to the non-compact case, because compact Lorentzian spaces are not causal. So the noncommutative version of gravitation is a theory with huge potential especially as an unification theory, but is still at this time a {\it mathematical} theory. The Lorentzian problem should be solved in order to make physical predictions and applications.

\end{document}